\documentclass[11pt,axodraw]{article}
\usepackage{amssymb}
\usepackage{amsmath}

\begin{document}

\title{Average Kinetic Energy of Heavy Quark ($\mu_\pi^2$) \\
inside Heavy Meson in $0^-$ State by Bethe-Salpeter Method } \vspace{4mm}

\author{C.~ S.~ Kim\footnote{cskim@yonsei.ac.kr}~~ and~~
Guo-Li~ Wang\footnote{glwang@cskim.yonsei.ac.kr} \\
{\it \small Department of Physics, Yonsei University, Seoul 120-749,
Korea}}
\date{}
\maketitle

\baselineskip=24pt
\begin{quotation}
\vspace*{1.5cm}
\begin{center}
  \begin{bf}
  ABSTRACT
  \end{bf}
\end{center}
\vspace*{0.5cm} \noindent The average kinetic energy of the heavy
quark inside $B$ or $D$ meson is computed by means of the
instantaneous Bethe-Salpeter method. We first solve the
relativistic Salpeter equation and obtain the relativistic wave
function and mass of $0^{-}$ state, then we use the relativistic
wave function to calculate the average kinetic energy of the heavy
quark inside heavy meson of $0^{-}$ state. We find that the
relativistic corrections to the average kinetic energy of the
heavy quark inside $B$ or $D$ meson are quite large and cannot be
ignored. We estimate  $\mu^2_\pi~ (= - \lambda_1)~ \approx~ 0.35~
(B^0,B^\pm)$, $~0.28~(D^0,D^\pm)$, $~ 0.43~(B_s)$, $~0.34~(D_s)$,
$~0.96~(B_c)$ and $0.62~(\eta_c)$ GeV$^2$.

\noindent
\end{quotation}

\newpage
 \setcounter{page}{1}
\section{Introduction}

In recent years, the study of hadronic processes involving heavy
quarks has attracted continuous interest both in experiment and in
theory. The difficulty of full theory of QCD, which is dynamic
theory describing the quark and gluon, lead us to the theoretical
achievements of the Heavy Quark Effective Theory (HQET)
\cite{review}. The latter describes the dynamics of heavy hadrons,
$i.e.$ hadrons containing a heavy quark $Q$, when $m_{_Q} \to
\infty$. The theory is based upon an effective lagrangian written
in terms of effective fields, which is a systematic expansion in
the inverse powers of the heavy quark mass $m_{_Q}$. The ${\cal
O}\Big({1 \over m_{_Q}}\Big)$ lagrangian reads as follows:
\begin{equation} {\cal L}={\bar h}_v i \; v \cdot D h_v+{1 \over 2 m_{_Q}}
{\bar h}_v [(iD_{\bot})^2] h_v +{g_s \over 2 m_{_Q}} {\bar h}_v
{\sigma_{\mu \nu} G^{\mu \nu} \over 2} h_v +{\cal O} \Big({1 \over
m_{_Q}^2}\Big)~,
\label{1}
\end{equation}
where the velocity-dependent field $h_v$ is the
heavy quark field, and $v_\mu$ is the heavy quark
four-velocity within the heavy
hadron. Then the total momentum is written as $p_{_Q}=m_{_Q}v+q$,
where the residual momentum $q$ is the difference between the
total momentum and the mechanical momentum; $ {D}^\mu=
{\partial}^\mu -i g A^\mu$ is the covariant derivative, and
${D_{\bot}}^\mu=D^{\mu}-v^{\mu}{v\cdot D}$ contains its components
perpendicular to the hadron velocity. In the hadron's rest frame
we have $(iD_{\bot})^2={\vec {D}}^2$. The second operator
appearing in Eq. (\ref{1}) corresponds to the kinetic energy
resulting from the residual motion of the heavy quark, and the third
one in Eq. (\ref{1})  the Pauli chromomagnetic interaction
operator which describes the interaction of the heavy quark spin
with the chromomagnetic gluon field. Their matrix elements can be
parameterized as follows \cite{afalk}:
\begin{equation}
\mu_\pi^2(H_Q)={< H_Q|{\bar h}_v ( {\vec {D}})^2
h_v| H_Q > \over 2 M_{H}}~,
\end{equation}
\begin{equation}\mu_G^2(H_Q)=
{< H_Q|{\bar h}_v {g\over 2}  \sigma_{\mu \nu} G^{\mu \nu} h_v|
H_Q > \over 2 M_{H}}~,
\end{equation}
where $H_Q$ denotes generically a hadron containing the heavy
quark $Q$ with the usual normalization $<H_Q|{\bar h}_v h_v|H_Q>=2
M_{H}$.

These two quantities are
interesting for several reasons. In the HQET, heavy
hadron mass is expected to scale with the heavy quark mass $m_{_Q}$
as:
\begin{equation}
M_{H}=m_{_Q}+{\bar \Lambda}+{\mu_\pi^2 - \mu_G^2
\over 2 m_{_Q}}+ ...~,
\end{equation}
where $\bar \Lambda$ represents the
difference between the mass of the hadron and that of the heavy
quark in  the $m_{_Q} \to \infty$ limit. In this
limit, it can be related to the trace anomaly  of QCD \cite{bsuv}:
$$
{\bar \Lambda}={1 \over 2 M_{H}}<H_Q| {\beta(\alpha_s) \over 4
\alpha_s} G^{\mu \nu} G_{\mu \nu}|H_Q>~,
$$
where $\beta$ is the Gell-Mann-Low function.
Moreover, if the inclusive semileptonic
width of a heavy hadron is calculated by an expansion in the
powers of ${1 \over m_{_Q}}$, the following results are found: the
leading term of the expansion coincides with the free quark decay
rate (spectator model); no corrections of order ${1 \over m_{_Q}}$
appear in the rate, and the ${1 \over m_{_Q}^2}$ corrections depend on
$\mu_\pi^2$ and $\mu_G^2$ \cite{bigi}. As a consequence, these
parameters enter in the ratio of hadron lifetimes and in the
lepton spectrum in inclusive transitions, which in principle are
quantities directly comparable with experimental data.
Many authors have given theoretical estimates of
${\mu_{\pi}}^2$ and ${\mu_G}^2$ using different methods, but
different results are obtained for the estimation of
$\mu_\pi^2$ (see Table 1). Even though there
may be different definitions of these two quantities,
our knowledge of them is still far from clear due to
large discrepancies, and a more careful study is
still needed.

\begin{table}[]\begin{center}
\caption{Theoretical estimates of the parameter $\mu^2_{\pi}$
of $B_{u,d}$
(QCDSR: QCD sum rules, HQSR: heavy-quark sum rules,  Exp.:
experimental data on inclusive decays, QM: quark models.)}
\vspace{0.5cm}
\begin{tabular}{|l|l|l|}\hline
Reference \rule[-0.25cm]{0cm}{0.7cm} & Method &
 $\mu^2_{\pi}$ [GeV$^2$] \\
\hline Eletsky, Shuryak \cite{ElSh} \hspace{-3mm}
\rule{0cm}{0.4cm} &
 QCDSR & $0.18\pm 0.06$ \\
Ball, Braun \cite{BaBr} & QCDSR & $0.52\pm 0.12$ \\
Neubert \cite{l1sr} & QCDSR & $0.10\pm 0.05$ \\[0.06cm]
\hline Gim\'enez $et~ al.$\ \cite{GiMS} \rule{0cm}{0.4cm} & Lattice &
 $-0.09\pm 0.14\!$ \\[0.06cm]
 A S Kronfeld $et~ al.$ \ \cite{kron}\rule{0cm}{0.4cm}& Lattice
  &$0.45\pm 0.12\!$ \\[0.06cm]
\hline Bigi $et~ al.$\ \cite{bsuv} \rule{0cm}{0.4cm} & HQSR & $>0.36$
 \\[0.06cm]
\hline Gremm $et~ al.$\ \cite{GKLW} \rule{0cm}{0.4cm} & Exp. &
 $0.19\pm 0.10$ \\
Falk $et~ al.$\ \cite{FLS2} & Exp. & $ 0.1\rightarrow 0.16$ \\
Chernyak \cite{Cher} & Exp. & $0.14\pm 0.03$ \\[0.06cm]
M Battaglia $et~ al.$\ \cite{mar} & Exp. & $ 0.17$ \\
\hline Hwang $et~ al.$\ \cite{Hwan} \rule{0cm}{0.4cm} & QM &
 $0.4\rightarrow0.6$ \\
De Fazio\ \cite{Fazi} & QM & $0.66\pm 0.13$ \\[0.06cm]
 S Simula\ \cite{simula} & QM    & $-0.089$   \\
 T Matsuki $et~ al.$\ \cite{mats}& QM & $0.238$ \\
\hline
\end{tabular}
\label{tab:kinetic}
\end{center}\end{table}

In this letter, we give a relativistically calculated version of
${\mu_{\pi}}^2$, $i.e.$  we calculate the average kinetic energy of the
heavy quark inside heavy meson in $0^-$ state by means of the Bethe-Salpeter
method \cite{BS}. We solve the relativistic Salpeter
equation \cite{salp} in Section 2, and give the mass and relativistic wave
functions of heavy meson in $0^{-}$ state in Section 3. Finally, we
use these relativistic wave functions to calculate the average kinetic energy
of the heavy quark in Section 4. Discussions and conclusions are also in
Section 4.

\section{Instantaneous Bethe-Salpeter Method}

It has been known that the Bethe-Salpeter (BS) equation is one of the
frameworks to describe bound state systems relativistically and has
a very solid basis in quantum field theory. So it is very often
used to describe bound state problems, and even in the current
literature many authors would like to base the constituent quark
model on the BS equation. For instance, in the constituent
quark model the mesons, corresponding quark-antiquark bound states,
are described by the BS equation as:
\begin{equation}
(\not\!{p_{_Q}}-m_{_Q})\chi(q)(\not\!{p_{q}}+m_{q})=
i\int\frac{d^{4}k}{(2\pi)^{4}}V(p,k,q)\chi(k)\;, \label{eq1}
\end{equation}
where $\chi(q)$ is the BS wave function with the total momentum
$p$ and relative momentum $q$, and $V(p,k,q)$ is the kernel
between the quarks in the bound state. The momenta $p_{_Q}, p_{q}$ are those of
the constituent quarks 1 and 2: For a
heavy meson with a heavy and a light valence quark, we can treat one of
these two constituents as a heavy quark and the other as a light
quark, $e.g.$ we treat the quark as the heavy quark $p_1=p_{_Q}$
and the anti-quark as the light quark $p_2=p_q$. The total
momentum $p$ and the relative momentum $q$ are defined as:
$$p_{_Q}={\alpha}_{1}p+q, \;\; {\alpha}_{1}=\frac{m_{_Q}}{m_{_Q}+m_{q}}~,$$
$$p_{q}={\alpha}_{2}p-q, \;\; {\alpha}_{2}=\frac{m_{q}}{m_{_Q}+m_{q}}~.$$
One can see that these definitions are just the same as in the HQET,
where ${\alpha}_{1}p$ is the mechanical momentum of the heavy
quark which describes  the heavy quark moving together with the
meson, and the relative momentum $q$ is nothing but the residual
momentum of the heavy quark inside meson. However, our method is not
the HQET and we do not have the limit of $m_{_Q} \to \infty$, so the
light quark momentum have the meaning analogous to that of the heavy quark.

The BS wave function $\chi(q)$ should satisfy the following
normalization condition:
\begin{equation}
\int\frac{d^{4}k d^{4}q} {(2\pi)^{4}}Tr\left[\overline\chi(k)
\frac{\partial}{\partial{p_{_0}}}\left[S_{1}^{-1}(p_{_Q})
S_{2}^{-1}(p_{q})\delta^{4}(k-q)+
V(p,k,q)\right]\chi(q)\right]=2ip_{_0}\;, \label{eq2}
\end{equation}
where $S_1(p_{_Q})$ and $S_2(p_{q})$ are the propagators of the two constituents.
In many applications, the kernel of the four-dimensional BS
equation is ``instantaneous'', $i.e.$ in the center of mass
frame of the concerned bound state ($\stackrel{\rightarrow}{p}=0$),
the kernel $V(p,k,q)$ of the
BS equation takes the simple form:
$$V(p,k,q) \Rightarrow V(k,q)=V(|\stackrel{\rightarrow}{k}|,
|\stackrel{\rightarrow}{q}|, \cos\theta)\;,$$
where $\theta$
is the angle between the vectors $\stackrel{\rightarrow}{k}$ and
$\stackrel{\rightarrow}{q}$. Then the BS equation may be reduced
to a three-dimensional one.
Compared with the conditions to solve a three-dimensional equation,
$i.e.$ to evaluate its eigenvalues and eigenfunctions, the
conditions to solve a four-dimensional one are much more complicated.
Thus if the kernel of the BS equation for the considered problem is
instantaneous, then we always would like to do the `reduction'
from four-dimensional to three-dimensional. Salpeter was the first
to do this reduction, so the reduced BS equation with
instantaneous kernel is also called the Salpeter equation. Here we
briefly repeat his method and solve the full Salpeter equation.
This equation is relativistic although it has an
instantaneous kernel, so we will obtain the relativistic wave
function of bound state.

Since in the HQET the heavy quark momentum is described by using the
covariant derivative $ {D}_\mu= {\partial}_\mu -i g  A_\mu$, and
the kinetic energy of the residual motion of the heavy quark by using a
covariant form ${D}_{\bot}$, it is convenient to write the BS
equation in a covariant form. To do this, we divide the relative
momentum $q$ into two parts, $q_{\parallel}$ and $q_{\perp}$, a
parallel part and an orthogonal one to the total momentum of the
bound state, respectively,
$$q^{\mu}=q^{\mu}_{\parallel}+q^{\mu}_{\perp}\;,$$
$$q^{\mu}_{\parallel}\equiv (p\cdot q/M^{2}_{H})p^{\mu}\;,\;\;\;
q^{\mu}_{\perp}\equiv q^{\mu}-q^{\mu}_{\parallel}\;.$$
Correspondingly, we have two Lorentz invariant variables:
\begin{center}
$q_{p}=\frac{(p\cdot q)}{M_{H}}\;, \;\;\;\;\;
q_{_T}=\sqrt{q^{2}_{p}-q^{2}}=\sqrt{-q^{2}_{\perp}}\;.$
\end{center}
In the center of mass frame $\stackrel{\rightarrow}{p}=0$, they turn out
to be the usual component $q_{0}$ and $|\stackrel{\rightarrow}{q}|$,
respectively. One can see that in the rest frame of bound state
the orthogonal residual momentum of the heavy quark is just the
orthogonal relative momentum, $i.e.$ $i\vec{D}=\vec{q}$.  Now the
volume element of the relative momentum $k$ can be written in an
invariant form:
\begin{equation}
d^{4}k=dk_{p}k^{2}_{T}dk_{_T}ds d\phi\;, \label{eq3}
\end{equation}
where $\phi$ is the azimuthal angle,
$s=(k_{p}q_{p}-k\cdot q)/(k_{_T}q_{_T})$.
The instantaneous interaction kernel can be
rewritten as:
\begin{equation}
V(|\stackrel{\rightarrow}{k}-\stackrel{\rightarrow}{q}|)=
V(k_{\perp},q_{\perp},s)\;. \label{eq4}
\end{equation}

Let us introduce the notations
$\varphi_{p}(q^{\mu}_{\perp})$ and $\eta(q^{\mu}_{\perp})$
for three dimensional wave function as follows:
$$
\varphi_{p}(q^{\mu}_{\perp})\equiv i\int
\frac{dq_{p}}{2\pi}\chi(q^{\mu}_{\parallel},q^{\mu}_{\perp})\;,
$$
\begin{equation}
\eta(q^{\mu}_{\perp})\equiv\int\frac{k^{2}_{_T}dk_{_T}ds}{(2\pi)^{2}}
V(k_{\perp},q_{\perp},s)\varphi_{p}(k^{\mu}_{\perp})\;.
\label{eq5}
\end{equation}
Then the BS equation can be rewritten as:
\begin{equation}
\chi(q_{\parallel},q_{\perp})=S_{1}(p_{_Q})\eta(q_{\perp})S_{2}(p_{q})\;.
\label{eq6}
\end{equation}
The propagators of the two constituents can be decomposed as:
\begin{equation}
S_{i}(p_{i})=\frac{\Lambda^{+}_{ip}(q_{\perp})}{J(i)q_{p}+\alpha_{i}M_{H}-\omega_{ip}+i\epsilon}+
\frac{\Lambda^{-}_{ip}(q_{\perp})}{J(i)q_{p}+\alpha_{i}M_{H}+\omega_{ip}-i\epsilon}\;,
\label{eq7}
\end{equation}
with
\begin{equation}
\omega_{ip}=\sqrt{m_{i}^{2}+q^{2}_{_T}}\;,\;\;\;
\Lambda^{\pm}_{ip}(q_{\perp})= \frac{1}{2\omega_{ip}}\left[
\frac{\not\!{p}}{M_{H}}\omega_{ip}\pm
J(i)(m_{i}+{\not\!q}_{\perp})\right]\;, \label{eq8}
\end{equation}
where $i=1, 2$ for heavy quark and light anti-quark, respectively,
$\omega_{1p}=\omega_{Q}$, $\omega_{2p}=\omega_{q}$, and
$J(i)=(-1)^{i+1}$. Here $\Lambda^{\pm}_{ip}(q_{\perp})$ satisfy
the relations:
\begin{equation}
\Lambda^{+}_{ip}(q_{\perp})+\Lambda^{-}_{ip}(q_{\perp})=\frac{\not\!{p}}{M_{H}}~,\;\;
\Lambda^{\pm}_{ip}(q_{\perp})\frac{\not\!{p}}{M_{H}}
\Lambda^{\pm}_{ip}(q_{\perp})=\Lambda^{\pm}_{ip}(q_{\perp})~,\;\;
\Lambda^{\pm}_{ip}(q_{\perp})\frac{\not\!{p}}{M_{H}}
\Lambda^{\mp}_{ip}(q_{\perp})=0~. \label{eq9}
\end{equation}
Due to these equations, $ \Lambda^{\pm}$ may be considered as
$p-$projection operators, while in the rest frame
$\overrightarrow{p}=0$ they turn to be the energy projection
operator.

Introducing the notations $\varphi^{\pm\pm}_{p}(q_{\perp})$ as:
\begin{equation}
\varphi^{\pm\pm}_{p}(q_{\perp})\equiv
\Lambda^{\pm}_{1p}(q_{\perp})
\frac{\not\!{p}}{M_{H}}\varphi_{p}(q_{\perp})
\frac{\not\!{p}}{M_{H}} \Lambda^{{\pm}}_{2p}(q_{\perp})\;,
\label{eq10}
\end{equation}
and taking into account $\frac{\not\!{p}}{M_{H}}\frac{\not\!{p}}{M_{H}}=1$, we
have
$$
\varphi_{p}(q_{\perp})=\varphi^{++}_{p}(q_{\perp})+
\varphi^{+-}_{p}(q_{\perp})+\varphi^{-+}_{p}(q_{\perp})
+\varphi^{--}_{p}(q_{\perp})
$$
With contour integration over $q_{p}$ on both sides of
Eq. (\ref{eq6}), we obtain:
$$
\varphi_{p}(q_{\perp})=\frac{
\Lambda^{+}_{1p}(q_{\perp})\eta_{p}(q_{\perp})\Lambda^{+}_{2p}(q_{\perp})}
{(M_{H}-\omega_{_Q}-\omega_{q})}- \frac{
\Lambda^{-}_{1p}(q_{\perp})\eta_{p}(q_{\perp})\Lambda^{-}_{2p}(q_{\perp})}
{(M_{H}+\omega_{_Q}+\omega_{q})}\;,
$$
and we may decompose it further into four equations as follows:
$$
(M_{H}-\omega_{_Q}-\omega_{q})\varphi^{++}_{p}(q_{\perp})=
\Lambda^{+}_{1p}(q_{\perp})\eta_{p}(q_{\perp})\Lambda^{+}_{2p}(q_{\perp})\;,
$$
$$(M_{H}+\omega_{_Q}+\omega_{q})\varphi^{--}_{p}(q_{\perp})=-
\Lambda^{-}_{1p}(q_{\perp})\eta_{p}(q_{\perp})\Lambda^{-}_{2p}(q_{\perp})\;,$$
\begin{equation}
\varphi^{+-}_{p}(q_{\perp})=\varphi^{-+}_{p}(q_{\perp})=0\;.
\label{eq11}
\end{equation}

In Ref. \cite{salp}, Salpeter considered the factor
${(M_{H}-\omega_{_Q}-\omega_{q})}$ being small, so he kept the
first of Eqs. (\ref{eq11}) only. It is the `original'
instantaneous approximation proposed by Salpeter and followed by
many authors in the literature. Whereas in this paper we
re-examine the BS equation with an instantaneous kernel, $i.e.$ we
try to deal with it exactly including the second of Eqs.
(\ref{eq11}). The complete normalization condition (keeping all
the four components appearing in Eqs. (\ref{eq11})) for BS equation
turns out to be:
\begin{equation}
\int\frac{q_{_T}^2dq_{_T}}{(2\pi)^2}tr\left[\overline\varphi^{++}
\frac{{/}\!\!\!
{p}}{M_{H}}\varphi^{++}\frac{{/}\!\!\!{p}}{M_{H}}-\overline\varphi^{--}
\frac{{/}\!\!\! {p}}{M_{H}}\varphi^{--}\frac{{/}\!\!\!
{p}}{M_{H}}\right]=2p_{_0}\;. \label{eq12}
\end{equation}
To solve the eigenvalue equation, one has to choose a definite
kernel of the quark and anti-quark in the bound state. As usual we
choose the Cornell potential, a linear scalar interaction
(confinement one) plus a vector interaction (single gluon
exchange):
\begin{equation}
I(r)=V_s(r)+V_0+\gamma_{_0}\otimes\gamma^0 V_v(r)= \lambda
r+V_0-\gamma_{_0}\otimes\gamma^0 \frac{4}{3}\frac{\alpha_s}{r}\;,
\label{eq13}
\end{equation}
where $\lambda$ is the string constant, $\alpha_s(r)$ is the
running coupling constant. Usually, in order to fit the data of
heavy quarkonia, a constant $V_0$ is often added to the scalar
confining potential.

It is clear that there exists infrared divergence in the Coulomb-like
potential. In order to avoid it, we introduce a factor
$e^{-\alpha r}$:
$$V_s(r)=\frac{\lambda}{\alpha}(1-e^{-\alpha r})~,$$
\begin{equation}
V_v(r)=-\frac{4}{3}\frac{\alpha_s}{r}e^{-\alpha r}\;. \label{eq14}
\end{equation}
It is easy to show that when $\alpha r\ll 1$, the potential becomes
identical with the original one. In the momentum space and
the rest frame of the bound state, the potential reads:
$$I(\stackrel{\rightarrow}{q})=V_s(\stackrel{\rightarrow}{q})
+\gamma_{_0}\otimes\gamma^0 V_v(\stackrel{\rightarrow}{q})~,$$
$$V_s(\stackrel{\rightarrow}{q})=-(\frac{\lambda}{\alpha}+V_0)
\delta^3(\stackrel{\rightarrow}{q})+\frac{\lambda}{\pi^2}
\frac{1}{{(\stackrel{\rightarrow}{q}}^2+{\alpha}^2)^2}~,$$
\begin{equation}
V_v(\stackrel{\rightarrow}{q})=-\frac{2}{3{\pi}^2}\frac{\alpha_s(
\stackrel{\rightarrow}{q})}{{(\stackrel{\rightarrow}{q}}^2+{\alpha}^2)}~.
\end{equation}
The coupling constant $\alpha_s(\stackrel{\rightarrow}{q})$ is
running:
$$\alpha_s(\stackrel{\rightarrow}{q})=\frac{12\pi}{27}\frac{1}
{\log (a+\frac{{\stackrel{\rightarrow}{q}}^2}{\Lambda^{2}_{QCD}})}~.$$
Here the constants $\lambda$, $\alpha$, $a$, $V_0$ and
$\Lambda_{QCD}$ are the parameters that characterize the
potential.

\section{Heavy Mesons in $0^{-}$ State}

Following the method \cite{changwang}, the general form for the
relativistic Salpeter wave function of the bound state
$J^{P}=0^{-}$ can be written as (in the center of mass system):
\begin{equation}
\varphi_{^{1}S_0}(\stackrel{\rightarrow}{q})=
M_{H}\left[{\gamma_{_0}}\varphi
_1(\stackrel{\rightarrow}{q})+\varphi_2(\stackrel{\rightarrow}{q})+
\frac{{\not\!q}_{\perp}}{M_{H}}
\varphi_3(\stackrel{\rightarrow}{q})+\frac{{\gamma_{_0}}{\not\!q}_{\perp}}{M_{H}}
\varphi_4(\stackrel{\rightarrow}{q})\right]\gamma_{_5}\;,
\end{equation}
where ${q}_{\perp}=(0,\stackrel{\rightarrow}{q})$, and $M_{H}$ is
the mass of the corresponding meson. The equations
$$\varphi^{+-}_{^{1}S_0}(\stackrel{\rightarrow}{q})
=\varphi^{-+}_{^{1}S_0}(\stackrel{\rightarrow}{q})=0$$ give the
constraints on the components of the wave function:
$$\varphi_3(\stackrel{\rightarrow}{q})=\frac{\varphi_2(\stackrel{\rightarrow}{q})
M_{H}(-\omega_{_Q}+\omega_{q})}{m_{q}\omega_{_Q}+m_{_Q}\omega_{q}}\;,
\;\;\;
{\varphi_4(\stackrel{\rightarrow}{q})}=-\frac{\varphi_1(\stackrel{\rightarrow}{q})M_{H}
(\omega_{_Q}+\omega_{q})}{m_{q}\omega_{_Q}+m_{_Q}\omega_{q}}\;.$$
Then we can rewrite the relativistic wave function of state
$0^{-}$ as:
\begin{equation}
\varphi_{^{1}S_0}(\stackrel{\rightarrow}{q})=M_{H}\left[
{\gamma_{_0}}\varphi_1(\stackrel{\rightarrow}{q})
+\varphi_2(\stackrel{\rightarrow}{q})-{\not\!q}_{\perp}\varphi_2(\stackrel{\rightarrow}{q})
\frac{(\omega_{_Q}-\omega_{q})}{(m_{q}\omega_{_Q}+m_{_Q}\omega_{q})}+
{\not\!q}_{\perp}\gamma_{_0}
\varphi_1(\stackrel{\rightarrow}{q})\frac{(\omega_{_Q}+
\omega_{q})}{(m_{q}\omega_{_Q}+m_{_Q}\omega_{q})}\right]\gamma_{_5}\;.
\end{equation}
{}From this wave function we can obtain the wave functions
corresponding to the positive and the negative projection,
respectively:
$$
\varphi^{++}_{^{1}S_0}(\stackrel{\rightarrow}{q})=
\frac{M_{H}}{2}\left[\left(\varphi_1(\stackrel{\rightarrow}{q})
+\varphi_2(\stackrel{\rightarrow}{q})\frac{\omega_{_Q}-\omega_{q}}{m_{_Q}-m_{q}}\right)\left(
\frac{m_{_Q}-m_{q}}{\omega_{_Q}-\omega_{q}}+{\gamma_{_0}}-\frac{{\not\!q}_{\perp}(m_{_Q}-m_{q})}
{m_{q}\omega_{_Q}+m_{_Q}\omega_{q}}\right)\right.$$\begin{equation}\left.+\frac{{\not\!q}_{\perp}\gamma_{_0}(\omega_{_Q}
+\omega_{q})}{(m_{q}\omega_{_Q}+m_{_Q}\omega_{q})}\left(\varphi_1(\stackrel{\rightarrow}{q})
+\varphi_2(\stackrel{\rightarrow}{q})\frac{m_{_Q}+m_{q}}{\omega_{_Q}+\omega_{q}}\right)\right]\gamma_{_5}\;,
\end{equation}
$$
\varphi^{--}_{^{1}S_0}(\stackrel{\rightarrow}{q})=
\frac{M_{H}}{2}\left[\left(-\varphi_1(\stackrel{\rightarrow}{q})
+\varphi_2(\stackrel{\rightarrow}{q})\frac{\omega_{_Q}-\omega_{q}}{m_{_Q}-m_{q}}\right)\left(
\frac{m_{_Q}-m_{q}}{\omega_{_Q}-\omega_{q}}-{\gamma_{_0}}-\frac{{\not\!q}_{\perp}(m_{_Q}-m_{q})}
{m_{q}\omega_{_Q}+m_{_Q}\omega_{q}}\right)\right.$$\begin{equation}\left.+\frac{{\not\!q}_{\perp}\gamma_{_0}(\omega_{_Q}
+\omega_{q})}{(m_{q}\omega_{_Q}+m_{_Q}\omega_{q})}\left(\varphi_1(\stackrel{\rightarrow}{q})
-\varphi_2(\stackrel{\rightarrow}{q})\frac{m_{_Q}+m_{q}}{\omega_{_Q}+\omega_{q}}\right)\right]\gamma_{_5}\;.
\end{equation}
And there are two more equations from the reduced BS equation (\ref{eq11}),
which will give
us coupled integral equations, and by solving them we  obtain the numerical
results for the mass and the wave function:
$$(M_{H}-\omega_{_Q}-\omega_
2)\left[\varphi_1({\stackrel{\rightarrow}{q}})+
\varphi_2({\stackrel{\rightarrow}{q}})\frac{\omega_{_Q}-\omega_{q}}{m_{_Q}-m_{q}}\right]=
-\int\frac{d{\stackrel{\rightarrow}{k}}}{(2\pi)^3}
\frac{1}{2\omega_{_Q}\omega_{q}(E_{Q}m_{q}+E_{q}m_{_Q})}$$
$$
\times\left\{(E_{Q}m_{q}+E_{q}m_{_Q})(V_s-V_v)\left[
\varphi_1({\stackrel{\rightarrow}{k}})(\omega_{_Q}\omega_{q}+m_{_Q}m_{q}-
{\stackrel{\rightarrow}{q}}^{2})+
\varphi_2({\stackrel{\rightarrow}{k}})(m_{q}\omega_{_Q}+m_{_Q}\omega_{q})
\right]\right.$$\begin{equation}\left.-(V_s+V_v)\left[
\varphi_1({\stackrel{\rightarrow}{k}})(m_{_Q}+m_{q})(E_{Q}+E_{q})+
\varphi_2({\stackrel{\rightarrow}{k}})(\omega_{_Q}-\omega_{q})(E_{Q}-E_{q})\right]
{\stackrel{\rightarrow}{q}} \cdot
{\stackrel{\rightarrow}{k}}\right\}~,
\end{equation}
$$(M_{H}+\omega_{_Q}+\omega_{q})\left[\varphi_1({\stackrel{\rightarrow}{q}})-
\varphi_2({\stackrel{\rightarrow}{q}})\frac{\omega_{_Q}-\omega_{q}}{m_{_Q}-m_{q}}\right]=
\int\frac{d{\stackrel{\rightarrow}{k}}}{(2\pi)^3}
\frac{1}{2\omega_{_Q}\omega_{q}(E_{Q}m_{q}+E_{q}m_{_Q})}$$
$$
\times\left\{(E_{Q}m_{q}+E_{q}m_{_Q})(V_s-V_v)\left[
\varphi_1({\stackrel{\rightarrow}{k}})(\omega_{_Q}\omega_{q}+m_{_Q}m_{q}-
{\stackrel{\rightarrow}{q}}^{2})-
\varphi_2({\stackrel{\rightarrow}{k}})(m_{q}\omega_{_Q}+m_{_Q}\omega_{q})
\right]\right.$$\begin{equation}\left.-(V_s+V_v)\left[
\varphi_1({\stackrel{\rightarrow}{k}})(m_{_Q}+m_{q})(E_{Q}+E_{q})-
\varphi_2({\stackrel{\rightarrow}{k}})(\omega_{_Q}-\omega_{q})(E_{Q}-E_{q})\right]
{\stackrel{\rightarrow}{q}} \cdot
{\stackrel{\rightarrow}{k}}\right\}~,
\end{equation}
where $E_{Q}=\sqrt{m_{_Q}^{2}+k^{2}_{_T}}$ and
$E_{q}=\sqrt{m_{q}^{2}+k^{2}_{_T}}$~. Finally the normalization
condition is
\begin{equation}\int\frac{d{\stackrel{\rightarrow}{q}}}{(2\pi)^3}4\varphi_1
({\stackrel{\rightarrow}{q}})\varphi_2({\stackrel{\rightarrow}{q}})M_{H}^2\left\{
\frac{\omega_{_Q}-\omega_{q}}{m_{_Q}-m_{q}}+\frac{m_{_Q}-m_{q}}{\omega_{_Q}-\omega_{q}}
+\frac{2{\stackrel{\rightarrow}{q}}^2(\omega_{_Q}m_{_Q}+\omega_{q}m_{q})}{(\omega_{_Q}m_{q}+\omega_{q}m_{_Q})^2}
\right\}=2M_{H}~.
\end{equation}

\section{Average Kinetic Energy of Heavy Quark inside Heavy Mesons in $0^{-}$ State}

\begin{table}[]\begin{center}
\caption{Three sets (1--3) of input parameters.
 $\lambda$ is in the unit of GeV$^2$, others are in the unit of GeV.}
\vspace{0.5cm}
\begin{tabular}
{|c|c|c|c|c|c|c|c|c|c|c|}\hline {\rm Set}&
$\alpha$&$V_0$&$\lambda$ &$\Lambda_{QCD}$&$m_b$ &$m_c$&$m_s$
&$m_d$&$m_u$
\\\hline
 (1)& 0.060&-0.60&0.20 &0.26&5.224 &1.7553&0.487 &0.311&0.305
\\\hline(2)&0.055 &-0.40&0.19 &0.24&5.130 &1.660&0.428 &0.285&0.278\\\hline
(3)&0.063 &-0.787&0.21 &0.275&5.3136&1.845 &0.557&0.352
&0.3465\\\hline
\end{tabular}
\end{center}
\end{table}

The average kinetic energy of the heavy quark inside heavy meson in
$0^{-}$ state, in the BS method, is proportional to the average spatial momentum squared:
\begin{equation}
\mu_{\pi}^{2}={\int\frac{d{\stackrel{\rightarrow}{q}}
 {\stackrel{\rightarrow}{q}}^2}{(2\pi)^3}2\varphi_{_1}
({\stackrel{\rightarrow}{q}})\varphi_2({\stackrel{\rightarrow}{q}})M_H\left\{
\frac{\omega_{_Q}-\omega_q}{m_{_Q}-m_q}+\frac{m_{_Q}-m_q}{\omega_{_Q}-\omega_q}
+\frac{2{\stackrel{\rightarrow}{q}}^2(\omega_{_Q} m_{_Q} +\omega_q
m_q)}{(\omega_{_Q} m_q+\omega_q m_{_Q})^2} \label{mupi}
\right\}}~.
\end{equation}
In order to solve numerically the relativistic Salpeter equation,
we use three different groups of input parameters
($i.e.$ parameters for the potential and the masses of quarks), as shown in Table 2, from
the best fit values \cite{para}:
\begin{center}
$a=e=2.7183$, $\alpha=0.06$ GeV, $V_0=-0.60$ GeV,
$\lambda=0.2$ GeV$^2$, $\Lambda_{QCD}=0.26$ GeV~~~ and \\
$m_b=5.224$ GeV, $m_c=1.7553$ GeV, $m_s=0.487$ GeV,
$m_d=0.311$ GeV, $m_u=0.305$ GeV.
\end{center}

With these three input parameter sets, we now solve the full Salpeter
equation and obtain the masses and wave functions of the ground
$0^{-}$ states. We list the calculated mass spectra of some
$0^{-}$ states as well as the measured experimental values in
Table 3. Then, by using the obtained wave function of heavy meson,
we  calculated $\mu_\pi^2$ from Eq. (\ref{mupi}), as shown in
Table 3.
\begin{table}[]\begin{center}
\caption{Mass spectra and $\mu^{2}_{\pi}$, for heavy mesons in
$0^{-}$ states with three sets (1--3) of input parameters. `Ex'
means the results from experiments \cite{PDG} and `ER' is the
error of experimental values. `Th' means the results from our
theoretical estimate.} \vspace{0.5cm}
\begin{tabular}
{|c|c|c|c|c|c|c|c|c|c|}\hline
&$B_c$&$B_s$&$B_d$&$B_u$ &$\eta_c$&$D_s$&$D_d$&$D_u$\\\hline $M$
GeV(Ex)
&6.4&5.3696&5.2794&5.2790&2.9797&1.9685&1.8693&1.8645\\\hline ER
of Ex &$\pm$0.4&$\pm$0.0024&$\pm$0.0005&0.0005
&$\pm$0.0015&$\pm$0.0006&$\pm$0.0005&$\pm$0.0005\\
\hline $M$ GeV(Th)(1)
&6.296&5.3654&5.2804&5.2778&2.9791&1.9688&1.8687&1.8655\\\hline
$M$ GeV(Th)(2)
&6.304&5.3670&5.2804&5.2762&2.9795&1.9691&1.8699&1.8650\\\hline
$M$ GeV(Th)(3)
&6.292&5.3656&5.2806&5.2788&2.9799&1.9690&1.8673&1.8650\\\hline
$\mu^{2}_{\pi}$
GeV$^2$(1)&0.955&0.434&0.348&0.345&0.615&0.339&0.280&0.277\\\hline
$\mu^{2}_{\pi}$
GeV$^2$(2)&0.906&0.429&0.354&0.350&0.596&0.331&0.280&0.277\\\hline
$\mu^{2}_{\pi}$
GeV$^2$(3)&0.958&0.446&0.350&0.347&0.636&0.352&0.286&0.284\\\hline
\end{tabular}
\end{center}
\end{table}
As can be seen from Table 3, if we change the values of the input
parameters (sets 1--3) used in solving the Salpeter equation, we
find that the obtained values of $\mu^{2}_{\pi}$ are almost
unchanged (especially for $B_d$, $B_u$, $D_d$ and $D_u$ mesons)
when these parameters give a reasonably good fit of mass spectra.
Therefore, we notice that our results for $\mu_\pi^2$ are quite
insensitive to the model parameters within the instantaneous BS
method. We also note that the average kinetic energies of the
heavy quark in different mesons differ significantly  even when
the heavy quark is the same, $e.g.$ the value of ${\mu_{\pi}}^2$
of the heavy quark is significantly larger in $B_s$ meson
($\approx 0.43$ GeV$^2$) than in $B_d$ ($\approx 0.35$ GeV$^2$) or
$B_u$ meson ($\approx 0.34$ GeV$^2$). The difference of about 0.08
GeV$^2$ is not a value which can be ignored compared with the
value of $\mu_\pi^2$ itself. The bigger value of  $\mu_\pi^2$
inside $B_s$ meson than inside $B_d$ or $B_u$ means that $b$ quark
has a smaller residual momentum in $B_d$ or $B_u$ than in $B_s$.
This implies  that $b$ quark is bounded more deeply in $B_d$ or
$B_u$ than in $B_s$ meson. In other words, the kinetic energy of
the same $b$ quark in heavy meson is more restrained by a light
partner quark than by a heavy one, which is consistent with the
running behavior of $\alpha_s$. Since our calculation of the
average kinetic energy of the heavy quark has used the
relativistic wave functions obtained from the full Salpeter
equation, our results of the average kinetic energy $\mu_\pi^2$
are relativistic. Note that our results are quite different from
the previously estimated ones of the potential model  \cite{Hwan,
Fazi, simula}. This shows that the relativistic corrections are
quite large, and cannot be ignored.

In Table 4, we also show the calculated theoretical uncertainties
for our results of the mass and average kinetic energy when we
allow variations of all the input parameters simultaneously within
5\% range of the central values. In comparison,  our result for
$B_{u,d}$
$$
\mu^2_\pi \approx 0.31 - 0.38~{\rm GeV}^2 , ~~~~ {\rm
[our~~estimate]}
$$
is a little larger than the recently  experimentally derived CLEO
values of
$$\mu^2_{\pi}=0.25\pm 0.05~,~~~ \cite{ex1}$$ and
$$\mu^2_{\pi}=0.24\pm 0.11~.~~~ \cite{ex2}$$

\begin{table}[]\begin{center}
\caption{The calculated uncertainties (in per cents)
if we allow changes of all input parameters simultaneously within
$ 5\%$ of the central values.
} \vspace{0.5cm}
\begin{tabular}
{|c|c|c|c|c|c|c|c|c|c|}\hline &$B_c$&$B_s$&$B_d$&$B_u$
&$\eta_c$&$D_s$&$D_d$&$D_u$\\\hline ${\Delta M}/M$ & $\pm 6.5$ &
$\pm 6.0$ & $\pm 5.8$ & $\pm 5.8$ & $\pm 7.2$ & $\pm 7.5$ & $\pm
7.3$ & $\pm 7.2$ \\\hline ${\Delta\mu^{2}_{\pi}}/\mu^{2}_{\pi}$ &
$\pm 7.0$ & $\pm 12.0$ & $\pm 10.6$ & $\pm 10.7$ & $\pm 9.5$ &
$\pm 10.3$ & $\pm 10.3$ & $\pm 10.5$  \\\hline
\end{tabular}
\end{center}
\end{table}

In conclusion, we calculated the average kinetic energy of the
heavy quark inside $B$ or $D$ meson by means of the instantaneous
Bethe-Salpeter method. We solved the relativistic Salpeter
equation and obtained the relativistic wave function and mass of
$0^{-}$ state. Then we used the relativistic wave function to
calculate the average kinetic energy of the heavy quark inside the
heavy $0^{-}$ state. We obtained $\mu^2_\pi~ (= - \lambda_1)~
\approx~ 0.35~ (B^0,B^\pm)$, $~0.28~(D^0,D^\pm)$, $~ 0.43~(B_s)$,
$~0.34~(D_s)$, $~0.96~(B_c)$ and $0.62~(\eta_c)$ GeV$^2$.
\\

\noindent We thank Chao-Hsi Chang, G. Cvetic and M. Olsson for
careful reading of the manuscript and their valuable comments. The
work of C.S.K. was supported in part by  CHEP-SRC Program and  in part
by Grant No. R02-2003-000-10050-0 from BRP of the KOSEF.
The work of G.W. was supported in part by BK21 Program and
in part by National Natural Science Foundation of China.
\\

\end{document}